\documentclass[aps,prd,groupedaddress,showpacs,nofootinbib,amssymb,balancelastpage,preprintnumbers,floatfix]{revtex4}

\usepackage[dvipdfmx]{graphicx}
\usepackage{graphicx,bm,color}

\usepackage{amsmath}
\usepackage{amssymb}
\usepackage{amsfonts}
\usepackage{physics}
\usepackage{cases}
\usepackage{cancel}
\usepackage{hyperref}
\usepackage[normalem]{ulem}
\usepackage{subfigure}
\usepackage{here}
\usepackage{color}
\usepackage{comment}

\usepackage{tikz}
\usetikzlibrary{matrix}

\allowdisplaybreaks[3]

\begin{document}
	

\title{ 
Walking-dilaton hybrid inflation with $B$ - $L$ Higgs embedded in dynamical scalegenesis   
}

\author{Jie Liu}\thanks{{\tt liujie22@mails.jlu.edu.cn}}
\affiliation{Center for Theoretical Physics and College of Physics, Jilin University, Changchun, 130012,
	China}

\author{He-Xu Zhang}\thanks{{\tt hxzhang18@163.com}}
\affiliation{Center for Theoretical Physics and College of Physics, Jilin University, Changchun, 130012,
	China}

\author{Hiroyuki Ishida}\thanks{{\tt ishidah@pu-toyama.ac.jp}}
\affiliation{Center for Liberal Arts and Sciences, Toyama Prefectural University, Toyama 939-0398, Japan}

\author{Shinya Matsuzaki}\thanks{{\tt synya@jlu.edu.cn}}
\affiliation{Center for Theoretical Physics and College of Physics, Jilin University, Changchun, 130012,
	China}%

\begin{abstract}
We propose a hybrid inflationary scenario based on eight-flavor hidden QCD with the hidden colored fermions being in part gauged under $U(1)_{B-L}$. This hidden QCD is  almost scale-invariant, so-called walking, and predicts the light scalar meson (the walking dilaton) associated with the spontaneous scale breaking, which develops 
the Coleman-Weinberg (CW) type potential as the consequence of the nonperturbative 
scale anomaly, hence plays the role of an inflaton of the small-field inflation. 
The $U(1)_{B-L}$ Higgs is coupled to the walking dilaton inflaton, which is dynamically induced from the so-called bosonic seesaw mechanism. 
We explore the hybrid inflation system involving the walking dilaton inflaton and the $U(1)_{B-L}$ Higgs as a waterfall field. 
We find that observed inflation parameters tightly constrain the $U(1)_{B-L}$  breaking 
scale as well as the walking dynamical scale to be $\sim 10^9$ GeV and $\sim 10^{14}$ GeV, respectively, so as to make the waterfall mechanism worked. 
The lightest walking pion mass is then predicted to be around 500 GeV. Phenomenological perspectives including embedding of the dynamical electroweak scalegenesis and 
possible impacts on the thermal leptogenesis are also addressed. 

\end{abstract}

\maketitle

\section{Introduction}

Inflation  
serves as solutions to several serious problems of the Standard Big Bang Cosmology,  such as horizon, flatness, and monopole 
problems, 
and also offers a robust framework for the generation of primordial density perturbations leading to the formation of large-scale structures in the universe that currently one sees. 
Among various inflationary models,  small field inflation (SFI) models based on the Coleman-Weinberg (CW) potential~\cite{Coleman:1973jx}  
%
has a potential merit also in particle physics perspective because possessing a possibility to make a nontrivial hotlink between the inflationary and the scale generation mechanisms, say, the dynamical origin of mass beyond the Standard Model (BSM) of particle physics.  
The CW-SFI coupled to some BSM 
was first explored long ago, back in 1980s, 
in a context such as the nonsupersymmetric Grand Unified Theories (GUTs)~\cite{Linde:1981mu,Albrecht:1982wi,Shafi:1983bd}. 
The ballpark of the CW-SFI  
has recently been extended and scaled down to 
the conformal extension of the SM including BSM sectors at 
as low as TeV scale, which can also have high sensitivity to the collider experimental probes, such as those at the LHC.  

In such a context, one attractive model in the CW-SFI would go with 
the scale-invariant extension of the SM with 
$U(1)_{B-L}$ gauge symmetry along with the $U(1)_{B-L}$ Higgs field in addition to the SM Higgs field~\cite{Okada:2016ssd}. There the 
$U(1)_{B-L}$ Higgs field develops the CW-type potential to  
play the role of the inflaton of the CW-SFI, and also accounts for the generation of 
the tiny neutrino mass scale by the seesaw mechanism. 
The dynamical realization of 
the electroweak symmetry breaking is also achieved via the (phenomenologically small enough) scale-invariant portal coupling between the $U(1)_{B-L}$ and the SM Higgs fields. 
Embedding the CW-SFI along with the electroweak symmetry 
breaking into QCD-like gauge theories has also been addressed in the literature~\cite{Iso:2014gka,Ishida:2019wkd,Cacciapaglia:2023kat,Zhang:2023acu}.   

However, the CW-SFI involves the intrinsic problem:  
to achieve a sufficient number of e-folds consistent with observations of the cosmic microwave background (CMB) fluctuations, 
the inflaton field is required to initiate the slow-roll (SR) from a region close to the false vacuum (the origin of the inflation potential), which is extremely distant from the true vacuum created by the quantum scale anomaly, but still needs to go somewhat away from the origin (the scale-symmetric phase). That is the initial place problem. 
Without an ad-hoc assumption, along the flat direction the CW-inflaton field would keep staying at the origin, 
in the symmetric phase, which can be stabilized by the possible thermal potential 
surrounded around the inflaton field made of the SM and/or BSM sector plasma. 
This state keeps going until the potential barrier between the false (the origin) and true vacua is gone: it is the end of the supercooling caused by the almost-scale 
invariant CW nature around the origin. Then, the inflaton would tunnel into 
a place in the other side of the barrier, close to the true vacuum (broken phase), and would start to SR to accumulate the e-folding number. 
Since the place where the inflaton tunnels out is thus governed by the scale-broken phase in the potential which is no longer almost flat, 
the distance of this SR journey is generically too short to account for 
the desired e-folding ($\sim 40 - 60$) supported from the presently observed universe 
via the CMBs. Thus, the initial place problem calls for dynamical trapping mechanism also taking into account the supercooling cosmological phase transition feature. 

Recently, dynamical trapping mechanisms have been discussed to resolve this initial place issue~\cite{Iso:2015wsf,Zhang:2023acu}. 
Some of the present authors, in particular, have proposed~\cite{Zhang:2023acu} the trapping mechanism intrinsic, closely tied to the supercooling feature of 
the cosmological CW-type phase transition: 
when the phase transition is ultra-supercooling in the presence of  
an explicit scale breaking effect (tadpole term) in the potential, the false vacuum place gets shifted away from the origin as temperature cools down.  
This mechanism allows the inflaton to dynamically start the SR down 
toward the true vacuum, 
due to the tadpole effect immediately after the potential barrier is gone -- which is also energetically favored over the tunneling process. 
In~\cite{Zhang:2023acu} 
the authors have explicitly employed a multiple scalar (meson) model underlined by many-flavor QCD, which develops the CW-type potential along the flat direction corresponding 
to the composite dilaton profile direction as was also addressed in~\cite{Ishida:2019wkd}.  
It has been demonstrated that the proposed trapping mechanism 
works consistently with all observational bounds on inflationary parameters, 
including realization of the large enough e-folding number. 



In~\cite{Ishida:2019wkd,Zhang:2023acu} the CW-SFI has proceeded along only the composite dilaton direction. The SM Higgs $(H)$ portal coupling to the dilaton ($\chi$) was introduced (dynamically induced) there, like the form $\sim \lambda_{H\chi} \chi^2 |H|^2$, which, however, turns out to be extremely small: $\lambda_{H \chi} \sim (v_{\rm H}/v_\chi)^2 \sim 10^{-26}$, where $v_\chi \sim 10^{15}$ GeV consistently 
with the successful SFI~\cite{Ishida:2019wkd,Zhang:2023acu}, and $v_{H}$ is the electroweak scale $ \simeq 246$ GeV. Therefore, the variation of the 
SM Higgs field $H$ during the SFI of the dilaton $\chi$ gives negligibly tiny correction to the dilaton-CW potential,   
all the way up until the dilaton reaches the true vacuum at $v_\chi \sim 10^{15}$ GeV. 
The $U(1)_{B-L}$ Higgs field ($\phi$)  
may be, however, allowed to take a sizable portal coupling to the dilaton field. 
It is then suspected to play a role of what is called the waterfall~\cite{Linde:1993cn}, 
which triggers the end of the SR inflation earlier than the case with 
the single dilaton inflaton. 
Consequently, the true vacuum place of the $U(1)_{B-L}$ Higgs field, i.e., the vacuum expectation value (VEV) $v_\phi$, might be highly constrained, which would be impact 
on modeling of underlying theories embedding the $B-L$ gauge symmetry, such as GUTs, 
or on the thermal leptogenesis~\cite{Fukugita:1986hr} along with the constrained heavy Majorana mass scale 
proportional to this $v_\phi$, or even further cosmological implications in the post inflationary epoch, like in  
the nonperturbative particle production of lepton and baryon number asymmetry 
via the preheating mechanism~\cite{Dolgov:1989us,Traschen:1990sw,Kofman:1994rk,Shtanov:1994ce,Kofman:1997yn} 
(for reviews, see e.g., \cite{Kofman:1997yn,Amin:2014eta,Lozanov:2019jxc}).

In this paper, we extend the CW-SFI scenario in~\cite{Ishida:2019wkd,Zhang:2023acu}
by incorporating the $U(1)_{B-L}$ gauge symmetry in part into the dynamical scalegenesis triggered by eight-flavor hidden QCD 
together with an elementary $U(1)_{B-L}$ Higgs field as a waterfall field. 
We explore the hybrid inflation system involving the walking dilaton inflaton and the waterfall $U(1)_{B-L}$ Higgs field.

QCD with three colors and eight Dirac fermions in the fundamental representation of $SU(3)$ has been supported by lattice simulations to be so-called 
a walking gauge theory definitely possessing the chiral broken phase~\cite{Aoki:2013xza,LSD:2014nmn,Hasenfratz:2014rna}, which is emergent as 
a remnant of the Caswell-Banks-Zaks infrared fixed
point~\cite{Caswell:1974gg,Banks:1981nn}. 
The scalar meson much lighter than other hadrons has been supported to be present also there~\cite{LatKMI:2014xoh,LatKMI:2016xxi,LatticeStrongDynamics:2018hun}, that is what we shall call the walking dilaton, which is identified as 
the pseudo Nambu-Goldstone boson associated with 
the spontaneous breaking (but explicitly broken at the same time) of 
approximate scale invariance in the walking gauge theory. 
The walking dilaton potential is generated by the nonperturbative 
scale anomaly driven by the chiral symmetry breaking,
and is fixed by the anomalous Ward-Takahashi identity for the scale symmetry~\cite{Matsuzaki:2015sya,Matsuzaki:2012vc,Matsuzaki:2013eva}, which generically takes the form of logarithmic CW type. 
The dilaton-tadpole term can also be present arising as the chiral and scale 
explicit breaking due to the hidden fermion mass term, 
which plays the important role in addressing the trapping mechanism~\cite{Zhang:2023acu}. 
Thus the walking dilaton can be an inflaton of the CW-SFI as has been 
discussed in the literature~\cite{Ishida:2019wkd,Cacciapaglia:2023kat,Zhang:2023acu}.

The walking-dilaton portal coupling to the $U(1)_{B-L}$ Higgs is dynamically 
generated via the so-called bosonic seesaw mechanism with 
the classical scale-invariance~\cite{Calmet:2002rf,Kim:2005qb,Haba:2005jq,Antipin:2014qva,Haba:2015lka,Haba:2015qbz,Ishida:2016ogu,Ishida:2016fbp,Haba:2017wwn,Haba:2017quk,Ishida:2017ehu,Ishida:2019gri,Ishida:2019wkd}. 
The sign of the portal coupling is unambiguously fixed to be negative~\cite{Ishida:2019gri}, 
which is essentially due to the scale invariance and kinematic nature of the scalar exchange as the repulsive force 
contribution between scalar probes. 
When the walking dilaton reaches the true vacuum after the CW-SFI is made ended by 
the waterfall, 
this negative Higgs portal gives the VEV ($v_\phi$) of the waterfall at 
its true vacuum.

We investigate the hybrid dynamics involving the walking dilaton and the $U(1)_{B-L}$ 
Higgs fields 
in detail. 
In the literature~\cite{Cacciapaglia:2023kat} another type of the walking-dilaton hybrid inflation has been investigated with the walking pions as the waterfall field. 
In the present work, the walking pions, arising as the pseudo Nambu-Goldstone boson 
associated with the spontaneous chiral breaking, 
do not come into play in the hybrid system, because 
we will employ in the chiral broken phase of the walking gauge theory. 
This is also the flat direction with $\pi=0$ along which  
the walking dilaton gets light enough just like a `scalon' a la 
Gildener-Weinberg (GW)~\cite{Gildener:1976ih}, as has been 
shown in the literature~\cite{Kikukawa:2007zk,Miura:2018dsy,Zhang:2023acu,Zhang:2024vpp}. 
It turns out that 
observed inflation parameters tightly constrain the $U(1)_{B-L}$ breaking   
scale, $v_\phi$, as well as the walking dynamical scale $v_\chi$ as $v_\phi \sim 10^9$ GeV and $v_\chi \sim 10^{14}$ GeV, respectively, so as to make the waterfall mechanism worked. 
We find another characteristic prediction to the lightest walking pion mass, 
which is tied with the size of the walking dilaton tadpole effect, i.e., 
the initial place of the SR. 
The walking pion mass is constrained to be in a tiny range around 500 GeV.

This paper is organized as follows. 
In Sec.~\ref{Sec2} we propose a model of dynamical scalegenesis 
based on the eight-flavor QCD with $U(1)_{B-L}$ gauged in part, and 
derive the walking dilaton potential and the portal coupling term with the 
$U(1)_{B-L}$ gauge Higgs, relevant to the hybrid inflation system that we will later employ.  
Section~\ref{sec3} provides the formulae for the inflation parameters, to be constrained by 
the current CMB observations. The initial condition of the walking dilaton inflaton 
and the way of the trapping mechanism with the thermal corrections are addressed  
as well. 
In Sec.~\ref{sec4} we discuss the constraints from the CMB observations and 
give predictions derived from the tightly bonded model parameter space 
in the present hybrid inflation system. 
Summary of the present paper is given in Sec.~\ref{summary}, where 
we also give several discussions related to the predictions from the 
present hybrid inflation scenario, in a view of phenomenological consequences in light of 
collider physics such as the LHC run 3 and prospects on the cosmological production 
of dark matter candidates predicted from the present model. 
Detailed supplements regarding computations of thermal corrections in the present model 
and the related discussions are given in Appendix~\ref{LSM}.



\section{The proposed model description and dynamical scalegenesis} 
\label{Sec2}

The hidden walking gauge theory with the $SU(3)_{HC}$ gauge symmetry, that we assume to be present in 
the inflationary epoch, includes  
eight flavor Dirac fermions, $\psi^{i}(i=1,\dots,8)$, 
which are (at this moment) electroweak ($SU(2)_W \times U(1)_Y$) and QCD ($SU(3)_c$) singlets, but are partly charged by the $U(1)_{B-L}$ gauge. 
To ensure the $U(1)_{B-L}$-gauge-anomaly-free in the SM fermion sector, we also introduce right-handed Majorana neutrinos 
$N_{R}^{(i=1,2,3)}$ forming three families. 
Adding the $U(1)_{B-L}$ Higgs $\phi$ and the SM Higgs doublet $H$, 
the charge assignments for one-family fermion contents 
under the $SU(3)_{\mathrm{HC}}\times U(1)_{B-L}\times SU(3)_{c}\times SU(2)_{W}\times U(1)_{Y}$ are as follows. 

\begin{table}[!ht]
\centering
\begin{tabular}{cccccc}
	 \hline \hline
	                & $SU(3)_\mathrm{HC}$ & $U(1)_{B-L}$ & $SU(3)_c$ & $SU(2)_W$ & $U(1)_Y$ \\ \hline
	 $\psi^1_{L/R}$ & 3                   & +1           & 1         & 1         & 0        \\
	 $\psi^2_{L/R}$ & 3                   & -1           & 1         & 1         & 0        \\
	 $\psi^3_{L/R}$ & 3                   &  0           & 1         & 1         & 0        \\
	 $\vdots$       & $\vdots$            &  $\vdots$    & $\vdots$  & $\vdots$  & $\vdots$ \\
	 $\psi^8_{L/R}$ & 3                   &  0           & 1         & 1         & 0        \\ \hline
	 $\phi$         & 1                   & +2           & 1         & 1         & 0        \\
	 $N_R$          & 1                   & -1           & 1         & 1         & 0        \\ \hline
	 $q_L$          & 1                   & 1/3          & 3         & 2         &  1/6     \\
	 $l_L$          & 1                   & -1           & 1         & 2         & -1/2     \\
	 $u_R$          & 1                   & 1/3          & 3         & 1         &  2/3     \\
	 $d_R$          & 1                   & 1/3          & 3         & 1         & -1/3     \\
	 $e_R$          & 1                   & -1           & 1         & 1         & -1       \\ \hline
	 $H$            & 1                   &  0           & 1         & 2         & 1/2      \\
	 \hline \hline
\end{tabular}
\end{table}

We assume the classical scale invariance at the Planck scale in a sense of the renormalization group evolution of all the sectors, so that no mass terms are 
generated for $H$ and $\phi$, until the dimensional transmutation is induced 
by the hidden walking QCD sector at the intrinsic scale $\Lambda_{\rm HC}$. 
This can be interpreted and realized (by ``self-tuning") as the renormalization condition including 
over-Planckian quantum gravity such as the asymptotically safe quantum gravity~\cite{Wetterich:2016uxm}. 
The present model description is similar to the one proposed in the literature~\cite{Ishida:2017ehu} in a sense of embedding the $U(1)_{B-L}$ gauge in the dynamical scalegenesis driven by hidden QCD, though the latter did not possess a walking nature for the hidden QCD sector.

The model Lagrangian thus consists of the scale-invariant SM sector with $U(1)_{B-L}$ gauged, and the hidden QCD sector which is hidden chiral invariant, 
and gauge-invariant Yukawa interactions involving the $U(1)_{B-L}$ Higgs $\phi$: 
\begin{equation}
	\mathcal{L}_y =  -y_{\psi} \bar\psi_1\psi_2\phi -\sum_{} y_N \bar N_R^{c} N_R\phi 
 +\mathrm{h.c.}
 \,,  \label{Ly}
\end{equation}
where the Yukawa coupling to $N_R$ is understood to be present for each generation of $N_R$ with the basis chosen so as to make the flavor mixing diagonalized. 
In addition, we assume the presence of the explicit chiral-scale breaking term for the hidden QCD sector. 
The form is taken to be like a current quark mass term as is present in ordinary QCD, in a $SU(8)_V$ symmetric manner, for simplicity: 
\begin{equation}
    \mathcal{L}_m 
    = -m_0 \sum_{i=1}^{8}\bar F_i F_i 
\,, \label{Lm}
\end{equation}
with $m_0$ being assumed to be so small that the approximate low-energy chiral dynamics works well as in ordinary QCD, where $F$ denotes the $SU(8)$ flavor multiplet 
as $F = (\psi^1, \cdots ,\psi^8)^{T}$.

At the intrinsic scale of hidden QCD, $\Lambda_\mathrm{HC}$,  
the hidden chiral symmetry is spontaneously broken down: $SU(8)_L \times SU(8)_R \to SU(8)_V$, which is triggered by the dynamical generation of  
hidden-fermion condensate $\langle\bar F_i F_j\rangle \sim \Lambda_\mathrm{HC}^3 \cdot \delta_{ij}$. This is also the essential source of the dynamical scale symmetry breaking which at the same timing also causes the nonperturbative scale anomaly via 
the Miransky scaling~\cite{Miransky:1984ef} (sometimes also called  Berezinsky-Kosterlitz-Thouless scaling) intrinsic to the conformal phase transition~\cite{Miransky:1996pd,Kaplan:2009kr}: 
$m_F \sim \Lambda_{\rm HC} e^{- \pi/\sqrt{\alpha/\alpha_c} -1}$, 
where $\alpha$ is the fine structure constant of hidden QCD 
and $\alpha_c$ denotes the critical coupling above which the chiral symmetry breaking
takes place. This scaling surely realizes a large scale hierarchy in the chiral broken phase during the walking regime, spanned by the infrared scale, $m_F$, 
and the ultraviolet scale, $\Lambda_{\rm HC}$, at which $m_F$ is generated. 
This Miransky scaling also induces the walking dilaton potential 
of CW type consistently with the anomalous Ward-Takahashi identity for the scale 
symmetry~\cite{Matsuzaki:2012vc,Matsuzaki:2013eva}, as will be seen right below.

At the critical coupling scale $\Lambda_{\rm HC}$, 
the walking dilaton is generated as a composite of the flavor singlet bilinear, $\bar{F}F$, and can be as light as or less than the dynamical mass scale of $F$, $m_F$, 
reflecting the pseudo Nambu-Goldstone boson for the spontaneous breaking of the scale symmetry~\cite{LatKMI:2014xoh,LatKMI:2016xxi,LatticeStrongDynamics:2018hun}. 
At the scales $\lesssim m_F$, the low-lying walking hadron spectra are thus 
governed by only the walking pseudos, i.e., the walking dilaton and sixty-three walking pions (up to the singlet pion, like the $\eta'$ state in ordinary QCD which 
gets as heavy as $\sim 4 \pi m_F$ due to the axial anomaly in the walking gauge sector).

We may therefore replace the hidden fermion-chiral bilinear $ \bar F_{Ri} F_{Lj} $ as 
\begin{equation}
 \bar F_{Ri} F_{Lj} \approx \langle \bar F_{R} F_{L} \rangle \cdot \left( \frac{\chi}{v_\chi}\right)^{3-\gamma_m} \cdot U_{ij}
\label{FF}
\end{equation}
where $U=e^{2i\pi/f_\pi}$ stands for the chiral field embedding the walking pion fields $\pi = \sum_{a=1}^{63} \pi^a \cdot T^a$ in the nonlinear realization, 
together with the pion decay constant $f_\pi$ associated with 
the spontaneous chiral breaking. The chiral condensate $\langle \bar F_{R} F_{L} \rangle = \langle \bar F_{L} F_{R} \rangle = \langle \bar F F \rangle$ is $SU(8)$ flavor universal and parity invariant, dictated per flavor, 
and develops the mass anomalous dimension $\gamma_m$, which is $\simeq 1$ during the walking regime 
as has been clarified in the context of the walking technicolor~\cite{Yamawaki:1985zg,Bando:1987we,Bando:1986bg}.

The explicit-chiral breaking mass term in Eq.(\ref{Lm}) induces the walking dilaton $\chi$ potential term like 
\begin{equation}
	-\frac{C}{16}\chi^a \mathrm{Tr}[U+U^\dagger] 
 \,. 
\label{tad}
\end{equation}
The introduced power parameter $a$ 
is read off from 
the overlap amplitude between the $F$-fermion (isosinglet) bilinear $\bar{F}F$ and the dilaton state as 
$
\langle 0| \bar{F}F |\phi \rangle 
$. 
If the $\bar{F}F$ state is simply saturated by single pole only in 
the low-energy limit, the low-energy theorem for the scale symmetry 
reads~\cite{Matsuzaki:2012vc,Matsuzaki:2013eva} 
$
\langle 0| \bar{F}F |\phi \rangle 
= (3-\gamma_m)/v_\phi \langle \bar{F}F \rangle 
$, i.e., one finds that $a = (3-\gamma_m)$. 
As noted in the literature~\cite{Ishida:2019wkd}, 
however, 
possible mixing of the two-quark state $\bar{F}F$ with other flavor-singlet 
scalars, like a glueball or tetraquark states, might be present including 
also the axial anomaly effect, as is the case with ordinary QCD. 
In that case 
the parameter $a$ in Eq.(\ref{tad}) would generically be undermined until fully solving the mixing structure, that is beyond the current scope.  
In the present work, as has been done in the literature~\cite{Ishida:2019wkd,Zhang:2023acu}, 
we take a conservative 
limit with $a=1$, 
so that the $\chi$ couples to the $\bar{F}F$ as if it were a conventional 
singlet-scalar component as seen in the linear sigma model~\cite{Ishida:2019wkd,Zhang:2023acu,Zhang:2024vpp}.    
The choice $a=1$ would be reasonable 
if the linear sigma model gives a good low-energy description 
for the underlying walking gauge theory in terms of 
the chiral-breaking structure. 
As has been discussed in the literature~\cite{Zhang:2023acu,Zhang:2024vpp}, 
this ansatz is indeed compatible with 
the flat direction description along the walking 
dilaton direction arising from the many-flavor linear sigma model 
based on the GW approach.

The walking dilaton thus develops the CW-type potential reflecting 
the nonperturbative scale anomaly part together with the explicit chiral-scale breaking part in Eq.(\ref{tad})~\cite{Matsuzaki:2012vc,Matsuzaki:2013eva,Miura:2018dsy,Ishida:2019wkd,Zhang:2023acu}, 
\begin{equation}
   V_\chi =  - C \cdot \chi + 
   \frac{\lambda_{\chi}}{4}\chi^4\left( \ln\frac{\chi}{v_\chi}+A\right) 
   \quad , \label{Vchi} 
\end{equation}
with 
\begin{align} 
\lambda_{\chi} &=\frac{16 N_c N_f}{\pi^4}\frac{m_F^4}{v_\chi^4}
\,, \notag\\ 
A &=-\frac{1}{4}+\frac{C}{\lambda_{\chi} v_\chi^3} 
\,, \notag \\ 
C &=\frac{N_c N_f m_\pi^2 m_F^2}{8 \pi^2 v_\chi} 
\,  \label{V-para}
\end{align}
(with $N_c=3$ and $N_f=8$),  
which are fixed by the stationary condition at the true vacuum $\chi=v_\chi$ and  
the scale anomaly matching~\cite{Matsuzaki:2013eva,Ishida:2019wkd} based on the anomalous Ward-Takahashi identity for the scale symmetry.  
Here $m_\pi$ in $C$ is the walking pion mass in the external gaugeless limit.

At the scales $\lesssim m_F$, the Yukawa coupling term between $U(1)_{B-L}$ Higgs $\phi$ field in Eq.(\ref{Ly}) generates the mixing between $\phi$ and 
and another composite scalar meson field $\Phi \sim \psi_1 \bar{\psi}_2$ (and its hermitian conjugate one), involving the walking dilaton in a scale-invariant manner: 
\begin{equation}
    \mathcal{L}_y \Bigg|_{\mu \lesssim m_F} \approx  - \chi^2\left[c_1 y_\psi(\phi^\dagger\Phi+\mathrm{h.c.})+c_2 |\Phi^2|\right]
\,.   
\end{equation}  
The coefficient $c_2$ is positive definite to ensure 
the $SU(8)_V$ symmetry at the true vacuum of the $SU(3)$ vectorlike gauge theory 
followed by the Vafa-Witten's theorem~\cite{Vafa:1983tf}. 
As noted at the beginning of the present section, our renormalization condition (i.e. the classical scale invariance) ensures no mass term for $\phi$ to be present until the scale goes down to $\Lambda_{\rm HC}$ when the dimensional transmutation is triggered by the hidden walking QCD. 
Thus the mixing structure between $\phi$ and $\Phi$ takes the seesaw form, that 
is called the bosonic seesaw~\cite{Calmet:2002rf,Kim:2005qb,Haba:2005jq,Antipin:2014qva,Haba:2015lka,Haba:2015qbz,Ishida:2016ogu,Ishida:2016fbp,Haba:2017wwn,Haba:2017quk,Ishida:2017ehu,Ishida:2019gri,Ishida:2019wkd}.  
Integrating out the heavier $\Phi$, we find the portal coupling between 
$\phi$ and $\chi$, 
\begin{equation}
    \mathcal{L}_y \Bigg|_{\mu < m_\Phi \sim m_F} 
    \approx -\lambda_{\phi\chi}|\phi|^2 \chi^2   
    \,,  \label{BS}
\end{equation}
where 
\begin{align} 
\lambda_{\phi\chi}=\frac{y_\psi^2c_1^2}{c_2}
\,.  
\end{align} 
Thus the negative portal coupling is dynamically generated by the bosonic seesaw mechanism~\cite{Ishida:2019gri}, so is the negative mass squared of $\phi$, 
\begin{equation}
    m^2_{\phi}=-\lambda_{\phi\chi} v^2_\chi
\,. \label{mphi}
\end{equation}
such that the VEV $v_\phi$ will spontaneously break 
the $U(1)_{B-L}$ gauge symmetry. 


Charging $\Psi=(\psi_3,\psi_4)^T$ to be $(2, 1/2)$ for $SU(2)_W \times U(1)_Y$,
additional Yukawa interactions between the SM Higgs doublet $H$, $\Psi$ and, say, 
$\psi_{5,6,7,8}$ in a way similar to the one for $\phi$ in Eq.(\ref{Ly}) can also give rise to the bosonic seesaw between $H$   
and the composite Higgs doublet $\Theta_i\sim \bar\Psi\psi_i$ $(i=5,...,8)$. 
It readily yields the negative mass square of the SM-like Higgs field via the 
negative Higgs portal coupling similar to the one for $\phi$ Eq.(\ref{BS}), 
as has extensively been discussed in the literature~\cite{Haba:2015lka,Haba:2015qbz,Ishida:2016ogu,Ishida:2016fbp,Haba:2017wwn,Haba:2017quk,Ishida:2017ehu,Ishida:2019gri,Ishida:2019wkd}. 
We can thus realize the electroweak symmetry breaking in the present framework of 
the dynamical scalegenesis, and will be back to this point with another look 
in the Summary and Discussion, later on.

With $\phi_1$ and $H$ VEVs at hand, 
the Yukawa terms of the right-hand neutrino fields lead to the mass matrix of 
seesaw form, 
\begin{equation}
	-\begin{pmatrix}
		\bar \nu_L & \bar N_R^C
	\end{pmatrix}
	 \begin{pmatrix}
		0       & y_\nu v_H     \\
		y_\nu v_H & y_N v_\phi 
	\end{pmatrix}
	\begin{pmatrix}
		\nu_L^C \\ 
		N_R
	\end{pmatrix}
 \,. \label{Eq:nu-seesaw}
\end{equation}
This yields the active neutrino mass as $m_\nu \sim \frac{y_\nu^2 v_H^2}{m_N}$ with $m_N = y_N v_\phi$.

Combining the $\phi-\chi$ portal interaction term in Eq.(\ref{BS}) and 
the quartic coupling term among $\phi$s with 
the walking dilaton potential term in Eq.(\ref{Vchi}), 
the total potential relevant to the hybrid inflation involving $\chi$ and the $B_L$ Higgs $\phi$ takes the form
\begin{equation}
    V(\phi,\chi)=
	-\lambda_{\phi\chi}|\phi|^2 \chi^2+\lambda_{\phi}(|\phi|^2)^2-C \cdot \chi +\frac{\lambda_{\chi}}{4}\chi^4\left( \ln\frac{\chi}{v_\chi}+A\right)+V_0
\,, \label{Vfull}
\end{equation}
where $V_0$ denotes the vacuum energy, which is $\simeq 24/\pi^4 m_F^4$ when 
the $\phi-\chi$ portal and the $\phi$ quartic terms are small enough, as will be the 
case with the present scenario.

\section{Walking dilaton hybrid inflation with $U(1)_{B-L}$ Higgs} 
\label{sec3}

Following the procedure in the literature~\cite{Miura:2018dsy,Zhang:2024vpp} based on the 
GW approach with the flat direction, 
the $\chi-\phi$ hybrid potential in Eq.(\ref{Vfull}) gets thermal corrections 
from the thermal plasma made of the walking mesonic sector,    
along the flat direction of $\chi \cos \theta$ with the mixing angle $\sin  \theta = v_\phi/v_\chi$, while $\phi$ does not develop the potential along the flat direction which is orthogonal to the $\phi$ direction. 
The $U(1)_{B-L}$ gauge-radiative corrections are assumed to be negligible compared to the walking mesonic loop ones, i.e., 
the $U(1)_{B-L}$ gauge coupling is required to satisfy $g_{B-L}^2\cdot (v_\phi/v_\chi)^2 \ll \lambda_\chi$, so that the dynamical trapping mechanism keeps intact as in the literature~\cite{Zhang:2023acu}. 
We will come back to the phenomenological consequences regarding this assumption, later in Summary and discussions. 
The false vacuum place of $\chi$ keeps shifted toward the true vacuum as 
temperature cools down, until the supercooling ends at $T=T_n$ as 
the signal of starting of the nucleation. 
The detailed computations of the thermal loop corrections are given in Appendix~\ref{LSM}. 

At $T>T_n$, 
the portal term ($-\lambda_{\phi\chi}\phi^2\chi^2$) in Eq.(\ref{Vfull}) which contributes as the negative curvature of $\phi$ can keep competing with the thermal mass term of $\chi$: 
		\begin{equation}
			(\alpha T^2-\lambda_{\phi\chi}\phi^2)\chi^2
   \,,  
		\end{equation}
  where the explicit value of $\alpha$ can be estimated by thermal loop computations, which can also be read off from the literature~\cite{Miura:2018dsy,Zhang:2024vpp} or Appendix~\ref{LSM}. 
  The thermal mass term will be dominated even in the presence of the portal coupling term, and  generate the barrier in the CW-type potential to trigger the (ultra)supercooling, which persists all the way up until the barrier is gone 
  at $T=T_n$. 
  Similarly, 
  even the $\phi$ direction is stabilized at the origin due to the thermal mass term, which is generated from and dominated by  
  the $\lambda_\phi$ interaction. 
  Thus $\phi$ keeps staying in the symmetric phase at $\phi=0$ and will never interfere the dynamical trapping mechanism, so that the SR inflation dynamically gets ready to start at $T=T_n$ in essentially the same manner as in~\cite{Zhang:2023acu}.

Thus the slow roll parameters ($\eta$ and $\epsilon$), the e-folding number $(N)$, the magnitude of the 
scalar perturbation $(\Delta_R^2)$, 
and the spectral index $n_s$  
are respectively defined through 
the zero-temperature potential $V$ in Eq.(\ref{Vfull}) as    
\begin{align}
\epsilon & =\epsilon_\phi+\epsilon_\chi=\frac{1}{2} M_{\mathrm{pl}}^2\left(\frac{V_\phi}{V}\right)^2+\frac{1}{2} M_{\mathrm{pl}}^2\left(\frac{V_\chi}{V}\right)^2\,, \nonumber \\
\eta_\phi & =M_{\mathrm{pl}}^2 \frac{V_{\phi \phi}}{V}\,, \nonumber \\
\eta_{\phi \chi} & =M_{\mathrm{pl}}^2 \frac{V_{\phi \chi}}{V}\,, \nonumber \\
\eta_\chi & =M_{\mathrm{pl}}^2 \frac{V_{\chi \chi}}{V}\,,\nonumber \\
N & =\frac{1}{M^2_{\mathrm{pl}}} \int_{\chi_{\mathrm{end}}}^{\chi_{\mathrm{ini}}} d \chi\left(\frac{V}{V_\chi}\right)\,,\nonumber \\
\Delta_{\mathrm{R}}^2 & =\frac{V}{24 \pi^2 M_{\mathrm{pl}}^4 \epsilon}\,,\nonumber \\
n_s & =1-6 \epsilon+2 \frac{\left(V_\phi\right)^2}{\left(V_\phi\right)^2+\left(V_\chi\right)^2} \eta_\phi+2 \frac{\left(V_\chi\right)^2}{\left(V_\phi\right)^2+\left(V_\chi\right)^2} \eta_\chi+4 \frac{V_\phi \cdot V_\chi}{\left(V_\phi\right)^2+\left(V_\chi\right)^2} \eta_{\phi \chi}
\,, \label{inf-para}
\end{align} 
where 
$M_{\rm pl}$ is the reduced Planck mass $\simeq 2.4 \times 10^{18}$ GeV, 
and the subscripts $\phi$ and $\chi$ attached on the potential $V$ denote
the functional derivative of $V$ with respective to $\phi$ and $\chi$, respectively. 
The initial place $\chi_{\rm ini}$ is dynamically determined by 
the trapping mechanism in~\cite{Zhang:2023acu} at $T=T_n$, while 
the endpoint $\chi_{\rm end}$ is fixed by the waterfall mechanism. 
Note that no thermal corrections are involved in the parameters in Eq.(\ref{inf-para}), because 
the SR gets started after the thermal barrier is gone, i.e., the thermal corrections get negligibly small.

In the case with $m_\pi \ll \chi \ll m_F \ll v_\chi$ and a small portal $
\lambda_{\phi \chi}$ coupling, which turns out to be naturally realized in the present model, 
we find that the $\eta_\chi$ and $\epsilon$ as well as 
the $\Delta_R^2$ and $N$ can have simpler parameter dependence~\cite{Ishida:2019wkd}  
\begin{align} 
\eta_\chi & \sim  \frac{M_{\rm pl}^2}{v_\chi^2} \frac{\chi^2}{v_\chi^2} \ln \frac{\chi^2}{v_\chi^2} 
\,, \notag \\ 
\epsilon & \sim \left( \frac{M_{\rm pl}}{v_\chi} \right)^2 
\left( \frac{m_\pi}{m_F} \right)^4 
\,, \notag \\ 
\Delta_R^2  & \sim 
\left( \frac{m_F}{v_\chi} \right)^4 \cdot \left( \frac{v_\chi}{M_{\rm pl}} \right)^6 
\left( \frac{m_F}{m_\pi} \right)^4 
\,, \notag \\ 
N & \sim 
\frac{(\chi_{\rm end} - \chi_{\rm ini})}{\sqrt{\epsilon} M_{\rm pl}} 
\sim  
\frac{(\chi_{\rm end} - \chi_{\rm ini}) v_\chi}{ M_{\rm pl}^2} 
\left( \frac{m_F}{m_\pi} \right)^2 
\,.  \label{approximations}
\end{align}

\begin{figure}[t]
	\centering
	\includegraphics[width=0.5\textwidth]{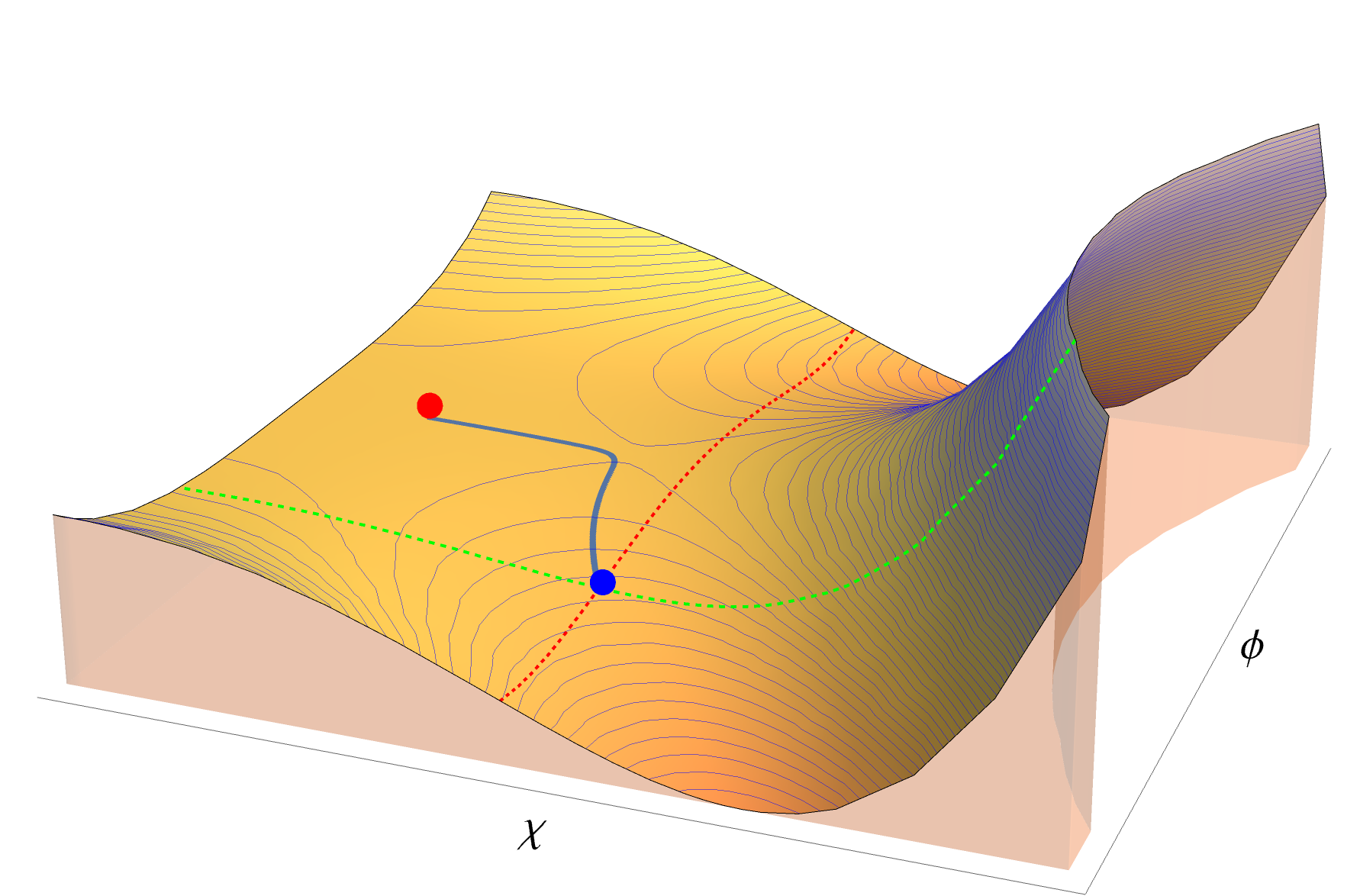}
	\caption{A schematic illustration of the present hybrid inflation in a view of the $\phi-\chi$ hybrid potential in Eq.(\ref{Vfull}). The blue curve indicates the trajectory of the SR inflation, which starts from the red dot point and ends up at the blue dot point. At the endpoint (blue dot), the slow-roll parameter $\eta_\phi$ along the waterfall $\phi$ direction reaches 1, $|\eta_{\phi}|\simeq 1$, while $\phi$ travels along the red dashed curve. At this point $\eta_{\chi}$ still keeps smaller than $1$ along the $\chi$ direction (green dashed curve). 
 Thus the waterfall field $\phi$ ends the SR epoch.}  
\label{inflation-plot}
\end{figure}

 A typical scenario goes like: at the nucleation temperature $T_n$, the SR inflation starts from the inflection point fixed by the trapping mechanism, which is seen on the surface of the hybrid potential along both dilaton and waterfall directions. 
 As the e-folding number $N$ gets accumulated in the SR epoch, the hybrid inflation should be stopped by the waterfall field $\phi$ when the slow-roll parameter $\eta_\phi$ reaches $1$ while other parameters still keep smaller than $\mathcal{O}(1)$. See Fig.~\ref{inflation-plot}, for a schematic picture of the hybrid inflation.

\section{Constraints and predictions}
\label{sec4}



The model parameter space is highly constrained by the observed data on cosmological parameters. First, we note that all the slow-roll parameters, 
should at least less than $\mathcal{O}(1)$ at the beginning of the SR inflation, $\chi=\chi_\mathrm{ini}$, which is identified as the pivot scale in the CMB observation. 
Since the $\chi$ direction around $\chi=\chi_{\rm ini}$ 
is almost flat due to the approximate scale invariance, 
the constraint can be essentially reflected only on $\eta_\phi$ at $\chi_{\rm ini}$, 
which can be evaluated from Eq.(\ref{inf-para}) as  
\begin{equation}
   | \eta_{\phi}|_{\chi=\chi_{\rm ini}} \approx M_{\mathrm{pl}}^2\frac{|V_{\phi\phi}|}{V_0} \Bigg|_{\chi=\chi_{\rm ini}}=\frac{M_{\mathrm{pl}}^2}{V_0}|-2\lambda_{\phi\chi}\chi_{\mathrm{ini}}^2+12\lambda_{\phi} \phi^2|<\mathcal{O}(1) \label{etaphi}
\,. 
\end{equation}

For $\phi$ to be an appropriate waterfall field, 
$\phi$ is not allowed to promptly get started to roll from the origin, otherwise the excessive 
e-folding number will be accumulated solely by the fast $\phi$ roll, spoiling the walking dilaton inflation. 
This requires a small enough portal coupling $\lambda_{\phi \chi}$, so that 
$\phi$ still stays around the origin, i.e., $\phi \sim 0$. 
As long as the $B-L$ dynamics is perturbative enough which is of our current concern,  
the quartic coupling $\lambda_\phi$ is at most of ${\cal O}(1 -10)$. 
In that case, the $\phi$ contribution in the second term of Eq.(\ref{etaphi}) can be neglected, 
because $\chi_{\rm ini}$ gets thermally shifted to relatively close to the true vacuum during the ultra-supercooling 
(See also the discussion later around Eq.(\ref{benchmark}) and Eq.(\ref{chi-ini})). 
Thus $\eta_\phi$ is adjusted to be small enough to be consistent with the observation as will be seen right below. 


For the present model to be consistent with the other cosmological parameters such as  e-folding number (predicted in a desired range), the scalar perturbation, and the spectral index must be like~\cite{ParticleDataGroup:2022pth} 
\begin{equation}
    N|_{\rm obs}\simeq 40 -  60, \quad 
    \Delta_\mathrm{R}^2|_{\rm obs} \simeq 2.137\times 10^{-9}, 
    \quad n_s|_{\rm obs} \simeq 0.968
\,. \label{obs}
\end{equation}
The tensor-to-scalar ratio $r (\simeq 16 \epsilon)$ is extremely small $(\simeq 10^{-25})$ 
in the case of the CW-SFI, as has been 
noted in the literature~\cite{Iso:2014gka,Ishida:2019wkd,Zhang:2023acu}, which is, combined with $n_s$ in Eq.(\ref{obs}), consistent with the current observation limit on the $n_s-r$ plane~\cite{ParticleDataGroup:2022pth}.  
Since the $\phi-\chi$ portal coupling is expected to be so small that 
the predictions solely from walking inflation dynamics in~\cite{Ishida:2019wkd,Zhang:2023acu} will not substantially be altered. 
This implies that a set of solutions to fit with the constraints on the inflation parameters in Eq.(\ref{obs}) can be present near the same reference point 
$(v_\chi, m_F) \sim (10^{15}\,{\rm GeV}, 10^{11}\,{\rm GeV})$ 
as was found in the literature. 
Thus we find a benchmark point near there,  
\begin{equation}
        v_\chi \simeq 5\times10^{14}\ \mathrm{GeV},\quad m_F \simeq 6\times10^{10}\ \mathrm{GeV},\quad m_\pi\simeq 500\ \mathrm{GeV}
\,, \label{benchmark}
\end{equation}
for $\lambda_\phi =  1$, 
with a small enough portal coupling 
$\lambda_{\phi\chi}\simeq (6 -2) \times 10^{-11}$.
The CW coupling of $\chi$ is estimated from $m_F$ and $v_\chi$ via Eq.(\ref{V-para}) 
as $\lambda_\chi \sim 4 \times 10^{-16}$. 
The walking dilaton mass at the true vacuum, approximately given as $M_\chi \simeq \sqrt{\lambda_\chi} v_\chi$ in the chiral and portal-less limit, is estimated to be $\simeq 1.4 \times 10^7$ GeV. 
Since the $U(1)_{B-L}$ gauge sector including the Higgs portal to the walking dilaton, as well as couplings to the SM sector, are negligibly small, the dynamical trapping mechanism 
in~\cite{Zhang:2023acu} keeps almost intact. 
The initial place of the SR inflation 
is thus estimated by solving the exactly the same bounce solution as in~\cite{Zhang:2023acu}, to be  
\begin{equation} 
    \chi_{\rm ini}(T_n)\sim 1.8\times 10^7\ \mathrm{GeV}
\,, \label{chi-ini}
\end{equation}
with $T_n\sim 4\times 10^5 \ \mathrm{GeV}$, which is compared to the critical temperature for the thermal-scale phase transition, $T_c \simeq 2 \times 10^{10}$ GeV.

As is evident from the approximate power-law scaling formulae in Eq.(\ref{approximations}), 
$\epsilon$, hence $\Delta_R^2$ and $N$ are highly sensitive to the variation of $m_\pi$. Thus $m_\pi$ is tightly constrained to be around $500\ \mathrm{GeV}$, as in Eq.(\ref{benchmark}), to be consistent with the CMB observations in Eq.(\ref{obs}).  
The $\phi-\chi$ portal coupling $\lambda_{\chi \phi}$ is also highly 
constrained to be as above, otherwise 

the SR inflation will be spoiled by the fast $\phi$ roll. 

Note that the predicted $m_\pi$ 
is much smaller than the one in the case with the single walking dilaton inflation~\cite{Ishida:2019wkd,Zhang:2023acu}, 
though the SR inflation dynamics is almost controlled by the walking dilaton even in the present case. 
Consequently, the benchmark parameter set in Eq.(\ref{benchmark}) 
is thus fairly stable, intrinsic, and almost unique to achieve the hybrid inflation.

Given the thus tightly fixed parameter set, 
the VEV of the $U(1)_{B-L}$ Higgs $\phi$ can be estimated 
by solving the stationary condition $\frac{\partial V}{\partial \phi}=0$ 
at $\phi=v_\phi,\chi=v_\chi$, to be 
\begin{equation}
    v_\phi \simeq 10^9\ \mathrm{GeV}, \qquad {\rm with} \qquad 
    m_\pi \simeq 500\ \mathrm{GeV}
\,. \label{pre}
\end{equation}
Thus the $U(1)_{B-L}$ breaking scale as well as the walking pion mass 
are firmly determined. 
This is the prediction arising as the consequence of 
the $U(1)_{B-L}$ waterfall coupled to the walking dilaton inflation, 
where both the initial and end points for the SR inflation 
are dynamically fixed.




\section{Summary and discussions}
\label{summary}

In summary, 
we have proposed a hybrid inflationary scenario embedded in the dynamical scalegenesis based on eight-flavor walking gauge theory with the hidden colored fermions being in part gauged under $U(1)_{B-L}$.  
We have identified the light scalar meson (the walking dilaton), associated with the spontaneous scale breaking develops, as the inflaton, which develops  
the Coleman-Weinberg (CW) type potential as the consequence of the nonperturbative scale anomaly in the walking gauge theory. 
The $U(1)_{B-L}$  Higgs is coupled to the walking dilaton SF inflaton, which is dynamically induced from the so-called bosonic seesaw mechanism, and develops  
the $U(1)_{B-L}$ breaking scale feeding the neutrino mass at the true vacuum. 
We have explored the hybrid inflation system involving the walking dilaton inflaton and the $U(1)_{B-L}$ Higgs as a waterfall field. 
We have found that observed inflation parameters tightly constrain the $U(1)_{B-L}$  breaking scale as well as the walking dynamical scale to be $\sim 10^9$ GeV and $\sim 10^{14}$ GeV, respectively (Eq.(\ref{benchmark})), so as to make the waterfall mechanism worked. 
The lightest walking pion mass is then predicted to be around 500 GeV (Eq.(\ref{pre})), which is also tied with the realization of the waterfall mechanism.

In closing, we give several comments on phenomenological and cosmological consequences regarding the prediction (Eq.(\ref{pre})).

The present scenario involves the $U(1)_{B-L}$ gauge sector with the $U(1)_{B-L}$ Higgs. As has been assumed, the $U(1)_{B-L}$ gauge contribution to the thermal and 
cosmological phase transition is negligible so that the walking dilaton inflation 
along with the dynamical trapping mechanism will substantially be intact as in the original literature~\cite{Zhang:2023acu}. The thus assumed $U(1)_{B-L}$ gauge coupling 
$g_{B-L}$ is constrained as $(g_{B-L} \cdot (v_\phi/v_\chi))^2 \lesssim \lambda_\chi$ 
(for more details, see also Appendix~\ref{LSM}). 
This implies $g_{B-L} \lesssim 10^{-3}$ and the $B-L$ gauge boson mass $m_X \lesssim 10^{6}$ GeV,  
for $v_\phi\sim 10^9$ GeV as listed in 
the benchmark point, in Eqs.(\ref{benchmark}) and (\ref{pre}). 
The current LHC searches for $Z'$ bosons of the so-called sequential SM 
in the dilepton, dijets, di$b$jets, and $t\bar{t}$ channels~\cite{ATLAS:2019erb,ATLAS:2020lks,ATLAS:2019fgd,CMS:2022eud,CMS:2019gwf,CMS:2021ctt} 
have placed upper bounds on the mass with the same couplings to SM fermions as the $Z$ boson's in the SM.  
The sequential SM $Z'$ boson has been ruled out up to the mass less than $\sim $ 5 TeV (most stringently placed from the dilepton channel) 
at around $137 - 139$ ${\rm fb}^{\rm -1}$ data with the center of mass energy $\sqrt{s} = 13$ TeV. 
Our $B-L$ gauge boson $X$ is much more weakly coupled to the SM quarks and leptons: $g_{B-L} \lesssim 10^{-5}$  
when its mass is within the LHC reach $\lesssim 10$ TeV. 
The $U(1)_{B-L}$ Higgs $\phi$ is too heavy ($\sim 10^9$ GeV) far above the mass scale of the LHC reach.  
Thus, the $U(1)_{B-L}$ sector in the present model is unlikely to be probed at the LHC run 3 or even the high-luminosity LHC with the luminosity of $\sim 3000 \,{\rm fb}^{-1}$.



Thus the presently introduced $U(1)_{B-L}$ gauge sector has no sensitivity to be probed at the LHC. When the electroweak symmetry breaking is embedded into the present dynamical scalegenesis, however, a 500 GeV walking ($B-L$ singlet) pion can be electroweakly produced at the LHC, and decay to di-weak bosons. 
As has been briefly noted below Eq.(\ref{mphi}), e.g., $\psi_{3,4}$ can form the electroweak doublet.  
Then a promising 500 GeV pion, which is neutral under the electroweak charge and free from the hidden QCD axial anomaly, is $\pi_{H}^0 \sim  \sum_{i=1}^7 \bar{\psi}_{i} \gamma_5 \psi_{i} - 7 \bar{\psi}_8 \gamma_5 \psi_8$: 
the $B-L$ and electroweak gauging is vectorlike (see the discussion in Sec.II) 
and the couplings to the walking pion fields arise as the commutator form 
$[V_{\rm gauge}^\mu, \pi_H]$, where 
\begin{align} 
V^\mu_{\rm gauge} \propto 
\left(
\begin{array}{ccc}  
X^\mu T_3 & 0_{2 \times 2} & 0_{2 \times 4} \\ 
0_{2 \times 2} & W^\mu_a T_a + B^\mu T_0  & 0_{2 \times 4} \\  
0_{4 \times 2} &  0_{4 \times 2} & 0_{4 \times 4} \\ 
\end{array} 
\right) 
\end{align} 
denotes the gauge field generator for the $B-L$ gauge field $X^\mu$, the $SU(2)$ gauge fields $W^\mu_{a=1,2,3}$,  
and $U(1)_Y$ gauge field $B^\mu$, with $T_{a=0, 1,2,3}$ being the Pauli matrices normalized as ${\rm tr}[T_aT_b] = \delta_{ab}/2$ where $T_0 = 1/2 \cdot 1_{2\times 2}$. 

The collider signals would then be like $\pi_H^0 \to WW, ZZ, Z \gamma$, and $\gamma\gamma$. This is phenomenologically identical to a fermiophobic axionlike particle (ALP) at the mass $\sim$ 500 GeV with the decay constant $f_\pi \sim m_F/4\pi \sim 10^{10}$ GeV. As seen from the analysis in the literature, say, ~\cite{Bauer:2017ris}, 
this sort of high mass ALPs with a large decay constant has no sensitivity to be probed at the LHC, simply due to too small coupling strengths to the electroweak gauge bosons and photon 
(corresponding to $C_{\gamma\gamma}^{\rm eff}/\Lambda \sim 10^{-8}\,{\rm TeV}^{-1}$ in terms of the coupling notation in the literature).  
The updated limits from the LHC on the ALP in the diphoton channel with the forward intact protons tagged have also been reported in~\cite{TOTEM:2021zxa} ($9.4\,{\rm fb}^{-1}$) and \cite{ATLAS:2023zfc} ($14.6\,{\rm fb}^{-1}$), which have placed the lower bound on the ALP decay constant, $f_\pi \gtrsim 25\,{\rm TeV}$ at the mass of 500 GeV. 
The current limit is still far lower than the benchmark point of our $\pi_H^0$.    
Thus, the presently predicted BSM sectors are unlikely to leave observable footprints in collider experiments, even at the high-luminosity LHC.

Other lightest walking pions with mass of 500 GeV, forming the 16-plet constructed only from the totally singlet hidden fermions $\psi_{5,6,7,8}$ 
(excluding the singlet representation which in part forms the $\eta'$-like state) and 
another $B-L$ chargeless one $\pi_H^{X0} \sim  \bar{\psi}_1 \gamma_5 \psi_1 -  \bar{\psi}_2 \gamma_5 \psi_2$ (associated with $T_3$ which gives the vanishing commutator with $X_\mu$),    
can be dark matters, because they are completely stable.
The cosmological abundance can be produced via the preheating mechanism~\cite{Dolgov:1989us,Traschen:1990sw,Kofman:1994rk,Shtanov:1994ce,Kofman:1997yn} 
(for reviews, see e.g., \cite{Kofman:1997yn,Amin:2014eta,Lozanov:2019jxc}) due to 
the nonadiabatic motion/oscillation of the walking dilaton in the post inflationary epoch.  
Since the walking dilaton strongly couples also to the electroweakly charged hidden fermions 
$\psi_3$ and $\psi_4$, hence $\pi_H^0$ above, 
the reheating would also be accessible at the same time as the dark matter production via the preheating. The detailed analysis would be noteworthy to be closely pursued in another publication.

From Eqs.~(\ref{Eq:nu-seesaw}) and (\ref{pre}), 
the largest neutrino Yukawa coupling is predicted to be of $\mathcal{O}(10^{-2.5})$ to explain the observed atmospheric neutrino mass scale with assuming $y_N$ to be unity.
In this case, one may consider the possibility of thermal leptogenesis~\cite{Fukugita:1986hr} with a mass of the heavy Majorana right-handed neutrino, $m_N \simeq 10^9~{\rm GeV}$. 
However, since our scenario has to require the reheating temperature, $T_R \lesssim T_n \sim 10^5~{\rm GeV}$ not to wash out the electroweak and $B-L$ breaking, 
the conventional thermal leptogenesis cannot be realized in the current setup. 
Instead, a class of the leptogenesis, called the low-scale leptogenesis, might still be possible to trigger. 
In this alternative case, the magnitudes of both $y_\nu$ in Eq.~(\ref{Eq:nu-seesaw}) and $y_N$ in Eq.~(\ref{Ly}) are interchanged and the predicted low-energy signatures would get highly  model-dependent. 
The explicit implications from our prediction to such model-dependent issues are beyond the current scope and would be performed elsewhere.

\section*{Acknowledgments} 

This work was supported in part by the National Science Foundation of China (NSFC) under Grant No.11747308, 11975108, 12047569, 
and the Seeds Funding of Jilin University (S.M.), 
and JSPS KAKENHI Grant Number 24K07023 (H.I.).

\appendix 

\section{Thermal loop computations based on the GW method} 
\label{LSM}

In this appendix, following the procedure in the literature~\cite{Zhang:2024vpp} 
we evaluate the thermal loop corrections to the walking dilaton potential in Eq.(\ref{Vfull}) 
by matching the one-loop effective potential arising from the scale-invariant 
linear sigma model along the flat direction with the walking dilaton potential.

\subsection{Mapping onto linear sigma model description with flat direction}

We start with writing the Lagrangian having the chiral $U(N_f)_L \times U(N_f)_R$ symmetry (with $N_f=8$ set finally) and scale invariance. 
The central building block is the linear-sigma model field, $M$, 
which forms an $N_f \times N_f$ matrix and transforms under the global chiral 
$U(N_f)_L \times U(N_f)_R$ symmetry as
\begin{equation}
	M \rightarrow g_L \cdot M  \cdot g^\dagger_R,\quad g_L,g_R \in U(N_f)\,, 
\end{equation}
and its hermitian conjugate $M^\dag$, 
where $g_L$ and $g_R$ stand for the transformation matrices belonging  
to the chiral-product group $U(N_f)_L \times U(N_f)_R$. 
In addition, the parity ($P$) and charge conjugate ($C$) 
invariance in the $M$ sector are assumed, under which $M$ transforms as 
$M \to M^\dag$ for $P$, and $M \to M^T$ for $C$. 

We include the $U(1)_{B-L}$ Higgs field $\phi$ coupled to this $M$ 
in a manner invariant under the global chiral 
$U(N_f)_L \times U(N_f)_R$ and the scale symmetry  
together with $C$ an $P$ invariance in the $M$ sector, 
The scale-invariant linear sigma model with $\phi$ is constructed as follows: 
\begin{equation}
	\mathcal{L} = \mathcal{L}_{B-L}\, + \mathrm{Tr} \left[\partial_\mu M^\dagger \partial^\mu M\right] - V(\phi,M)\,,
\label{LSM-L}
\end{equation}
where $\mathcal{L}_{B-L}$ is the $B-L$ sector Lagrangian including the canonical 
covariantized kinetic term of $\phi$, with $D_\mu \phi = (\partial_\mu - i g_{B-L} X_\mu) \phi$ , and the kinetic term of the $U(1)_{B-L}$ gauge field $X_\mu$. We have dropped 
the $U(1)_{B-L}$ gauge coupling to a part of $M$ (arising from couplings to mesons composed of $\psi_1$ and $\psi_2$), because those do not come into play at the one-loop effective potential 
of the walking dilaton (See also the discussion below Eq.(\ref{Pis})). 
In Eq.(\ref{LSM-L}) $V(\phi,M)$ denotes the scale-invariant potential which takes the form 
\begin{equation}
	\label{potential}
	V(\phi,M) = \lambda_1 \left(\mathrm{Tr}[M^\dagger M]\right)^2 + \lambda_2 \mathrm{Tr}\left[(M^\dagger M)^2\right] + \lambda_{\rm mix} |\phi|^2 \mathrm{Tr}[M^\dagger M] + \lambda_\phi |\phi|^4
\,, 
\end{equation} 
with $\lambda_h$, $\lambda_1$, 
and $\lambda_2$ being positive definite, while $\lambda_{\rm mix}$ (eventually) gives a negative portal coupling between $\chi$ (which is equivalent to $\sigma$ in Eq.(\ref{matrixfield}) with other meson fields set to zero) and $\phi$, 
after integrating out heavier meson fields, as has been noted in the main text.

The chiral $U(N_f)_L \times U(N_f)_R$ symmetry is assumed to be spontaneously broken down to the diagonal subgroup $U(N_f)_V$, reflecting the underlying vectorlike gauge theory. 
Thus $M$ and $\phi$ fields are parametrized as 
\begin{equation}
	\langle M \rangle =\frac{\bar{\sigma}}{\sqrt{2 N_f} }\cdot \,\mathbb{I}_{N_f \times N_f} \,,\quad \langle \phi \rangle=\frac{1}{\sqrt{2}} \bar{\phi} \,, 
\end{equation}
where $\mathbb{I}_{N_f \times N_f}$ is the $N_f$ by $N_f$ unit matrix. 
In terms of the background fields $\bar{\sigma}$ and $\bar{\phi}$, 
the tree-level potential is thus read off from Eq.(\ref{potential}) as 
\begin{equation}
	\label{treelevelV}
	V_{\rm tree} = \frac{1}{4} \left( \lambda_1 + \frac{\lambda_2}{N_f} \right)\bar{\sigma}^4 + \frac{\lambda_{\rm mix}}{4} \bar{\phi}^2 \bar{\sigma}^2 + \frac{\lambda_\phi}{4} \bar{\phi}^4\,.
\end{equation} 
To this potential, the potential stability condition requires   
\begin{equation}
	\lambda_\phi \ge 0, \qquad \lambda_{\rm mix}^2\le 4\left(\lambda_1+\frac{\lambda_2}{N_f}\right)\lambda_\phi\,. 
\label{stab}
\end{equation}

We apply the Gildener-Weinberg approach~\cite{Gildener:1976ih} 
and find the flat direction, which can be oriented along 
$\bar{\phi} \propto \chi$ and $\bar{\sigma} \propto \chi$ with 
the unified single background $\chi$. 
This proportionality to $\chi$ is clarified by solving mixing between $\bar{\phi}$ and $\bar{\sigma}$ arising 
in $V_{\rm tree}$ of Eq.(\ref{treelevelV}) with the mixing angle $\theta$, in such a way that  
\begin{align}
\bar{\phi} &= \chi \sin\theta \,, \notag\\ 
\bar{\sigma} & = \chi\cos\theta    
\,. \label{h-chi}
\end{align} 
Using this we rewrite the tree-level potential in Eq.(\ref{treelevelV}) 
as a function of $\chi$, to get 
\begin{equation}
	V_{\rm tree} = \frac{\chi^4}{4}\left[ \left( \lambda_1 + \frac{\lambda_2}{N_f} \right)\cos^4\theta + \lambda_{\rm mix} \cos^2\theta\sin^2\theta + \lambda_\phi \sin^4\theta\right]. 
\end{equation}
The flat direction condition, which requires $V_{\rm tree}$ to vanish and stationary along that direction, yields
\begin{equation}
	\tan^2\theta = \frac{-\lambda_{\rm mix}}{2\lambda_\phi}\,, \quad \quad 
	\lambda_{\rm mix}^2 = 4 \left( \lambda_1 + \frac{\lambda_2}{N_f} \right)\lambda_\phi\, , 
\label{FD}
\end{equation}
at certain renormalization group scale $\mu$. 
The negative $\lambda_{\rm mix}$ is required to be compatible with 
the flat direction condition together with the tree-level potential stability condition in Eq.(\ref{stab}).  This negativeness is consistent with the current dynamical scalegenesis, and the bosonic seesaw mechanism as noted in the main text.

Around the VEV in the flat direction $M$ and $\phi$ can be expanded as
\begin{equation}
	\label{matrixfield}
	M=\frac{\bar{\sigma}+\sigma+i\eta}{\sqrt{2N_f}}\cdot \mathbb{I}_{N_f \times N_f}+\sum_{a=1}^{N_{f}^{2}-1}\left(\xi^a+i \pi^{a}\right) T^{a}, \quad  \phi = \frac{1}{\sqrt{2}} (\bar{\phi}+\tilde{\phi}) \,,
\end{equation}
where $T_a$ are the generators of $SU(N_f)$ group in the fundamental representation and normalized as
${\rm Tr}(T^aT^b)=\delta^{ab}/2$, and $\tilde{h}$ denotes the Higgs fluctuation field.   
In Eq.(\ref{matrixfield}) $\sigma$ and $\eta$ are the hidden QCD isospin-singlet scalar and 
pseudoscalar fields, while $\xi^a$ and $\pi^a$ the hidden QCD isospin-adjoint scalar and pseudoscalar fields, respectively.  
These hidden QCD-sector fields would be regarded as mesons in terms of the expected underlying QCD-like gauge theory.  
The field-dependent mass-squares for $\tilde{\phi}$, $\sigma$, $\eta$, $\xi^a$, and $\pi^a$ 
then read  
\begin{align}
m_{\sigma}^2(\chi) 
&= 3 \left( \lambda_1+\frac{\lambda_2}{N_f} \right)\chi^2\cos^2\theta + \frac{\lambda_{\rm mix}}{2} \chi^2 \sin^2\theta = 2 \left( \lambda_1+\frac{\lambda_2}{N_f} \right)\chi^2\cos^2\theta = \frac{\lambda_{\rm mix}^2}{2 \lambda_\phi} 
\chi^2\cos^2\theta 
,, \notag \\
m_{\xi^a}^2(\chi) &= \left(\lambda_1 +\frac{3\lambda_2}{N_f}\right)\chi^2\cos^2\theta + \frac{\lambda_{\rm mix}}{2} \chi^2 \sin^2\theta =\frac{2\lambda_2}{N_f}\chi^2\cos^2\theta\,, \notag \\
m_\eta(\chi) &= m_{\pi^a}^2(\chi) = 0\,, \notag\\ 
m_{\tilde{\phi}}^2(\chi) &= -\lambda_{\rm mix} \chi^2\cos^2\theta\,,  
\label{FDM}
\end{align}
where we have used the flat direction condition in Eq.(\ref{FD}). 
Note that the Nambu-Goldstone bosons $\pi^a$ and $\eta$ 
associated with the spontaneous chiral breaking are surely massless along the flat direction.  

The angle $\theta$ defined in Eq.(\ref{h-chi}) simultaneously diagonalizes 
the $\phi\mathchar`-\chi$ mixing mass matrix, 
\begin{equation}
	\mathcal{M}^2=\begin{pmatrix}
		2\lambda_\phi v_\phi^2 &  \lambda_{\rm mix}v_\phi v_\sigma \\
		\lambda_{\rm mix}v_\phi v_{\sigma} &\displaystyle 2 \left(\lambda_1+\frac{\lambda_2}{N_f} \right) v_{\sigma}^2
	\end{pmatrix}\,, 
\end{equation}
in such a way that 
\begin{equation}
	\binom{\phi_1}{\phi_2}= \begin{pmatrix}
		\cos\theta &  -\sin\theta\\
		\sin\theta & \cos\theta
	\end{pmatrix} \binom{\tilde{\phi}}{\sigma}\,,  
\end{equation}
with the mass eigenstate fields $\phi_1$ and $\phi_2$. 
This eigenvalue system gives the tree level mass eigenvalues for 
the mass eigenstates $\phi_1$ and $\phi_2$ as  
\begin{equation}
	m_{\phi_1}^2=-\lambda_{\rm mix}v_\chi^2 \quad , \quad m_{\phi_2}^2=0\,.
\label{lambda-mix}
\end{equation}
At this moment, $\phi_2$ thus becomes massless (called the scalon~\cite{Gildener:1976ih}) having the profile along the flat direction. At the one-loop level, this $\phi_2$ acquires a mass as the flat direction is lifted by the quantum corrections, and becomes what is called the pseudo-dilaton due to the radiative scale symmetry breaking. On the other hand, $\phi_1$ has the profile  perpendicular to the flat direction, which does not develop its mass along the flat direction.

\subsection{Gildener-Weinberg type scalegenesis and thermal corrections}

Along the flat direction in Eq.(\ref{FD}), 
we compute the one-loop effective potential at zero temperature in the $\overline{\rm MS}$ scheme. We take the Landau gauge for the $U(1)_{B-L}$ gauge loop contributions. In this case the Nambu-Goldstone boson loop contributions are field-independent at the leading order in the resummed perturbation theory, so that they are decoupled in the effective potential analysis. 
The thermal corrections are then incorporated in an appropriate way at the consistent one-loop level.

The resultant one-loop effective potential at zero temperature takes the form   
\begin{equation}
	V_1(\chi) = A \chi^4 +B \chi^4 \log \frac{\chi^2}{\mu_{\rm GW}^2}\,,
\end{equation}
with
\begin{align}
	A =& \frac{\cos^4\theta}{16\pi^2}\Bigg[\left( \lambda_1+\frac{\lambda_2}{N_f} \right)^2 \left(\log \left( 2 \left( \lambda_1+\frac{\lambda_2}{N_f} \right) \cos^2\theta\right)-\frac{3}{2}\right) + (N_f^2-1)\frac{\lambda_2^2}{N_f^2}\left(\log \left( \frac{2\lambda_2}{N_f}\cos^2\theta\right)-\frac{3}{2}\right) 
 \notag \\
	&+\frac{\lambda_{\rm mix}^2}{4}\left(\log \left(|\lambda_{\rm mix}|\cos^2\theta\right)-\frac{3}{2}\right)\Bigg]
	+\frac{1}{64\pi^2 v_\chi^4}\sum_{i=X}    
 n_i m_i^4\left(\log \frac{m_i^2}{v_\chi^2}-c_i\right), 
 \notag \\
	B =&\frac{\cos^4\theta}{16\pi^2}\left[ \left( \lambda_1+\frac{\lambda_2}{N_f} \right)^2 + \left( N_f^2-1 \right)\frac{\lambda_2^2}{N_f^2}+\frac{\lambda_{\rm mix}^2}{4}\right] 
 + \frac{1}{64\pi^2 v_\chi^4}\sum_{i=X} 
 n_i m_i^4\,,
\label{A-B}
\end{align}
where 
$c_i=\frac{1}{2}\,(\frac{3}{2})$ for the transverse (longitudinal) polarization of the $U(1)_{B-L}$ gauge boson. 
We have assumed the RHMnus $N_R^{(1,2,3)}$ to be decoupled from the thermal plasma. 
The numbers of degree of freedom (d.o.f.) $n_i$ for 
$X$ is 3, and the mass 
can be written as $m_i^2(\chi)=m_i^2\frac{\chi^2}{v_\chi^2}$, with $m_i = g_{B-L} v_\phi/2$.   

The nonzero VEV of $\chi$ is associated with the renormalization scale $\mu_{\rm GW}$ via the stationary condition (as the consequence of the dimensional transmutation):
\begin{equation}
	\frac{\partial V_1(\chi)}{\partial \chi}=0 \quad \Rightarrow \quad \mu_{\rm GW} = v_\chi \exp \left( \frac{A}{2B}+\frac{1}{4} \right)\,.
\end{equation}
Correspondingly, the effective potential can be rewritten as
\begin{equation}
	V_1(\chi) = B \chi^4 \left(\log \frac{\chi^2}{v_\chi^2}-\frac{1}{2}\right)+V_0\,,\quad V_0=\frac{B v_\chi^4}{2} 
 \simeq \frac{\lambda_2^2 v_\chi^4}{32 \pi^2} \frac{N_f^2-1}{N_f^2}
 \,, \label{V1}
\end{equation}
where we have assumed $(g_{B-L} v_\phi/v_\chi)^2 \lesssim \lambda_{\rm mix}, \lambda_2$, 
so that the $U(1)_{B-L}$ loop contributions can be dropped in evaluating $B$, 
as has been done also in the main text. 
From $V_1(\chi)$ above,  
the radiatively generated mass of $\chi$ is also obtained as 
\begin{equation}
	M_\chi^2= \left.\frac{\partial^2 V_1(\chi)}{\partial \chi^2}\right|_{\chi = v_\chi} = 8 B v_\chi^2\,.
\label{Mchi}
\end{equation} 
In Eq.(\ref{V1}) $V_0$ denotes the vacuum energy, which is determined by the normalization condition $V_1(v_\chi)=0$, and the last approximation has been made by taking into account the flat direction condition 
Comparing Eq.(\ref{V1}) with the nonperturbative scale anomaly part in Eq.(\ref{Vfull}) , proportional to $\lambda_\chi$, we find the matching 
condition along the identical flat direction~\cite{Miura:2018dsy,Zhang:2023acu} 
\begin{align}
    \lambda_2^2 &=\frac{2\pi^2 N_f^2}{N_f^2-1}\lambda_\chi  
\,. 
\end{align}


We have ignored the $C$ term contribution in Eq.(\ref{Vfull}). 
This prescription turns out to work well, as long as  
the flat direction can still approximately work, which 
will not substantially alter the characteristic features on the cosmological phase transition in the present model.

By following the standard procedure, 
the one-loop thermal corrections are evaluated as 
\begin{align}
	V_{1,T}(\chi,T) = &\frac{T^4 }{2\pi^2} J_B\left(\frac{m_\sigma^2(\chi)}{T^2}\right) +  \frac{\left(N_f^2-1\right)T^4 }{2\pi^2} J_B\left(\frac{m_{\xi^i}^2(\chi)}{T^2}\right) + \frac{T^4 }{2\pi^2} J_B\left(\frac{m_{\tilde{\phi}}^2(\chi)}{T^2}\right) \notag \\
	 & +\frac{T^4 }{2\pi^2} \left[\sum_{i=X}  n_i J_{B}\left(\frac{m_i^2(\chi)}{T^2}\right)\right]\,, \label{V1T}
\end{align}
with the bosonic thermal loop function 
\begin{equation}
	J_{B}(y^2) = \int_{0}^{\infty}\mathrm{d}t\, t^2\ln\left(1 - e^{-\sqrt{t^2+y^2}}\right)\,.
\end{equation}
It has been shown that the perturbative expansion will break down since in the high-temperature limit higher loop contributions can grow as large as the tree-level and one-loop terms~\cite{Curtin:2016urg,Senaha:2020mop}. 
To improve the validity of the perturbation, we adopt the truncated full dressing  resummation procedure~\cite{Curtin:2016urg}, which is performed by the replacement $m_{i}^2(\chi)\rightarrow m_{i}^2(\chi) +\Pi_i(T)$ in the full effective potential. 
The thermal masses $\Pi_i(T)$ are computed as follows: 
\begin{align}
	&\Pi_{\sigma/\xi^i}(T) = \frac{T^2}{6}\left[ \left( N_f^2+1 \right) \lambda_1+2N_f\lambda_2+\frac{\lambda_{\rm mix}}{4}\right],\quad
	\Pi_\phi(T) = T^2\left(\frac{\lambda_\phi}{4} +\frac{g_{B-L}^2}{4}+\frac{\lambda_{\rm mix}}{24} +\frac{N_f^2}{12}\lambda_{\rm mix}\right), \notag\\
	& \Pi_X^L(T) = \frac{11}{6} g_{B-L}^2 T^2,\quad \Pi_X^T(T) = 0\,.
\label{Pis}
\end{align}

In general, the contributions from the daisy resummation are less important due to the fact that the phase transition completes well below the critical temperature in the supercooling case. 
In the benchmark point of the main text, given around Eqs.(\ref{benchmark}) and (\ref{pre}), from the field-dependent mass formulae in Eq.(\ref{FDM}) we read $m_{\tilde{\sigma}}^2 \sim (\lambda_{\chi \phi} v_\chi)^2 \sim (10^{3}\,{\rm GeV})^2$, 
and 
$|m_{\tilde{\phi}}|^2 \sim (\sqrt{\lambda_{\chi \phi}} v_\chi)^2 \sim (10^8 \, {\rm GeV})^2$, 
while $\Pi_{\sigma}(T_n) \sim (\sqrt{64/6 \lambda_{\rm mix}} T_n)^2 \sim (1 \,{\rm GeV})^2$ and $\Pi_\phi(T) \sim (g_{B-L} T_n)^2 \lesssim ( 10^2 \,{\rm GeV})^2$, 
where we have assumed $g_{B-L} (v_\phi/v_\chi) \lesssim \lambda_\chi$, so that 
the $U(1)_{B-L}$ gauge loop contributions are negligible in $V_1(\chi)$, in Eq.(\ref{V1}).  
Thus, indeed the daisy resummation contribution can safely be neglected when 
the cosmological phase transition takes place around $T=T_n \sim 10^5$ GeV.


\end{document}